\documentstyle[graphics]{mn}
\newif\ifAMStwofonts
\input psfig.sty

\def\lmc{LMC~X--2}
\def\sco{Sco~X--1}
\def\msun{${\rm M}_{\odot}$}
\def\rsun{${\rm R}_{\odot}$}
\def\ergsec{\thinspace\hbox{$\hbox{erg}\thinspace\hbox{s}^{-1}$}}

\title{Correlated Optical and X-ray Variability in LMC~X--2}
\author[K.E. McGowan et al.]
       {K.E. McGowan$^{1,2}$\thanks{email: mcgowan@lanl.gov}, P.A. 
Charles$^{2,3}$, D. O'Donoghue$^{4}$, A.P. Smale$^{5}$\\        
$^{1}$Los Alamos National Laboratory, MS D436, Los Alamos, NM 87545, USA\\
$^{2}$Department of Physics, University of Oxford, Oxford OX1 3RH\\
$^{3}$Department of Physics \& Astronomy, University of Southampton, 
Southampton S017 1BJ\\
$^{4}$South African Astronomical Observatory, PO Box 9, Observatory 7935, Cape 
Town, South Africa\\
$^{5}$Laboratory for High Energy Astrophysics, Code 662, NASA/Goddard Space 
Flight Center, Greenbelt, MD 20771, USA\\}

\pagerange{\pageref{firstpage}--\pageref{lastpage}}
\pubyear{2003}

\begin{document}

\maketitle

\label{firstpage}

\begin{abstract}
We have obtained high time resolution (seconds) photometry of \lmc\ in
December 1997, simultaneously with the Rossi X-ray Timing Explorer
({\em RXTE}), in order to search for correlated X-ray and optical
variability on timescales from seconds to hours.  We find that the
optical and X-ray data are correlated only when the source is in a
high, active X-ray state.  Our analysis shows evidence for the X-ray
emission leading the optical with a mean delay of $<20$~s.  The
timescale for the lag can be reconciled with disc reprocessing, driven
by the higher energy X-rays, only by considering the lower limit for
the delay.  The results are compared with a similar analysis of
archival data of \sco. 
\end{abstract}


\section{Introduction}

\lmc\ is the most X-ray luminous low mass X-ray binary (LMXB) known.
It was  first observed in early satellite flights \cite{leong71} and 
observations showed it to vary from L$_{\rm X}
\sim0.6$--$3\times10^{38}$\ergsec \cite{mark75,john79,long81}.  Using
a precise X-ray location Johnston et al.\ (1979) \lmc\ was optically
identified as a faint, $V\sim18.8$, blue star \cite{pak78,pak79}.  
 
X-ray light curves from {\em EXOSAT} \cite{bonn89} showed that the 
source was most variable in the highest energy range (3.6--11~keV), the 
variability decreased with energy and it was almost constant in the lowest 
energy range (0.9--2.4~keV).  \lmc\ displayed flaring activity which was 
characterized by a spectral hardening above an energy of $\sim$3.6~keV.  

The optical spectrum is that of a typical LMXB with weak H$\alpha$, 
H$\beta$ and He{\scriptsize II} $\lambda$4686 emission superimposed on
a blue continuum.  The characteristics of the optical spectrum, the
relatively soft X-ray spectrum, and the high X-ray to optical
luminosity (L$_{\rm X}$/L$_{opt} \sim600$), imply \lmc\ is similar to
galactic LMXBs (e.g. van Paradijs 1983).  The optical spectrum lacks
the Bowen blend, but this is probably due to the lower metal
abundances in the LMC \cite{john79}, which is also used to
account for the exceptionally high X-ray luminosities of the LMC X-ray
binaries \cite{motch89}.  \lmc's similarity to LMXBs in the
Galaxy suggested a likely short orbital period (i.e.\ $\la1$~d). 

However, in spite of a number of studies, the period of \lmc\ remains
uncertain.  Motch et al.\ (1985) and Bonnet-Bidaud et al.\ (1989)
found  evidence for a period of $\sim6.4$~h, whereas Callanan et al.\
(1990) found a periodicity of 8.15~h, and Crampton et al.\ (1990)
suggested a much longer period of $\sim12.5$~days.  The only previous
simultaneous optical and X-ray coverage of \lmc\ was very short (6~h)
and showed no correlation between the two light curves \cite{bonn89}.
Here we present the results of much more extensive simultaneous
optical and X-ray photometry of \lmc\ from December 1997, the aim of
which was to search for correlated X-ray and optical variability on
timescales from seconds to hours and to investigate the previously
claimed periodicities. 

\section{Observations and Data Analysis}

Observations of the optical counterpart of \lmc\ were performed using
the UCT--CCD fast photometer \cite{odon95} at the Cassegrain focus
of the 1.9~m telescope at SAAO, Sutherland on 1997 December 4--6.  The
UCT--CCD fast photometer is a Wright Camera $576\times420$ coated GEC
CCD, which was used half-masked so as to operate in frame-transfer
mode.  In this configuration, only half of the chip is exposed, and at
the end of the integration the signal is read out through the masked
half.  In this way, it is possible to obtain consecutive exposures of
as short as 1~s with no deadtime.  White light high-speed photometry
runs were carried out with integration times of 2--10~s (see Table
\ref{tab:saao_log}). 

\begin{table}
{\caption{\label{tab:saao_log} Log of high-speed photometry observations of 
\lmc\ : SAAO 1.9 m.}}
\begin{center}
\begin{tabular}{c|c|c}
\hline
{Start Time} & {Duration} & {Exposure Time} \\ 
{JD +2450000} & (hr) & (s) \\
\hline 
787.306 & 0.70 & 2 \\
787.355 & 1.35 & 2 \\
787.425 & 0.87 & 5 \\
787.510 & 0.95 & 2 \\
787.558 & 0.58 & 2 \\
788.306 & 0.23 & 2 \\
788.423 & 1.35 & 10 \\
788.490 & 0.57 & 10 \\
789.286 & 0.45 & 2 \\
789.315 & 0.57 & 2 \\
789.422 & 1.42 & 2 \\
789.489 & 0.53 & 2 \\
789.558 & 0.27 & 2 \\
\hline
\end{tabular}
\end{center}
\end{table}

As the seeing fluctuated during the observations, point spread
function (PSF) fitting was essential to obtain good photometry.  The
reductions were performed with the {\sc iraf} implementation of {\sc
daophot ii} \cite{stet87}.  For our purposes only the relative
brightness of a star is of importance, and so differential photometry
was applied to help reduce the effects of any variations in
transparency, employing 2 bright local standards within our field of
view.  This resulted in a relative precision of $\pm0.04$~mag per
frame. 

The X-ray data were obtained using the proportional counter array
(PCA) instrument on the {\em RXTE} satellite between 1997 December 2
18:52 UT and December 7 1:40 UT, in an observing strategy designed to
achieve the maximum amount of simultaneous coverage with the optical
photometry.  Data from all proportional counter unit (PCU) layers and
detectors were included in the creation of X-ray light curves in the
2--10~keV energy range.  We chose a time resolution of 1~s, rather
than the standard 16~s, for cross-correlation purposes.  Background
subtraction was performed utilising standard models generated by the
{\em RXTE}/PCA team. Further information about the X-ray observations
can be found in Smale \& Kuulkers (2000).  

\begin{figure*}
\resizebox*{0.9\textwidth}{.42\textheight}{\rotatebox{-90}{\includegraphics{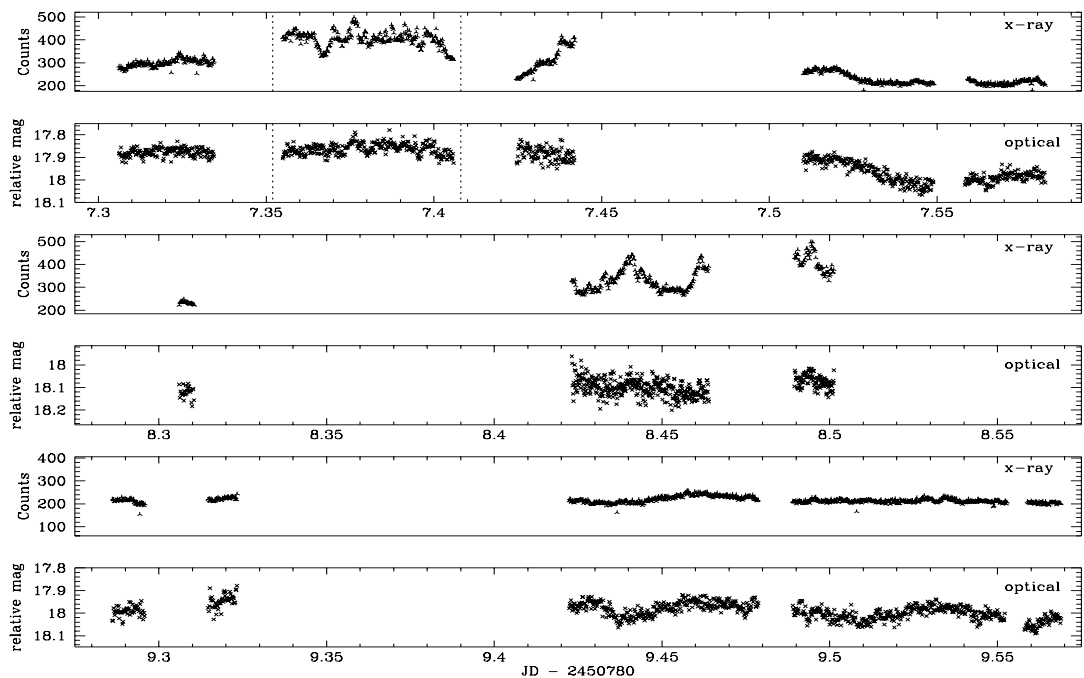}}}
\caption{All simultaneous SAAO optical (white light) and {\em
RXTE}/PCA X-ray data of \lmc\ taken over the period 1997 December
4--6.  The vertical dotted lines indicate the run in which correlated 
variability is found (see Section 3 and Fig.~\ref{fig:r303}).}\label{fig:opt_xr}
\end{figure*}

Optical data simultaneous with the {\em RXTE} data were obtained
covering 6.6~h, 4.68~h, and 6.8~h respectively on the three optical
observing nights (see Fig.~\ref{fig:opt_xr}).  The timing of the X-ray
data was measured in Julian Date (Terrestrial Time) [JD(TT)].  The
optical timing was measured in JD(UT) and had to be corrected for the 
accumulated leap seconds to produce timings in JD(TT) (see XTE Time
Tutorial\footnote{http://legacy.gsfc.nasa.gov/docs/xte/abc/time\_tutorial.html}). 

Although the UCT--CCD running in frame transfer mode can take
exposures as fast as 1~s, we found that the computer used to store the
images had a limit on how fast it could transfer the data.  Hence, the
effective exposure time for the shortest observations is 2.019~s.

\begin{figure*}
\resizebox*{0.85\textwidth}{.3\textheight}{\rotatebox{-90}{\includegraphics{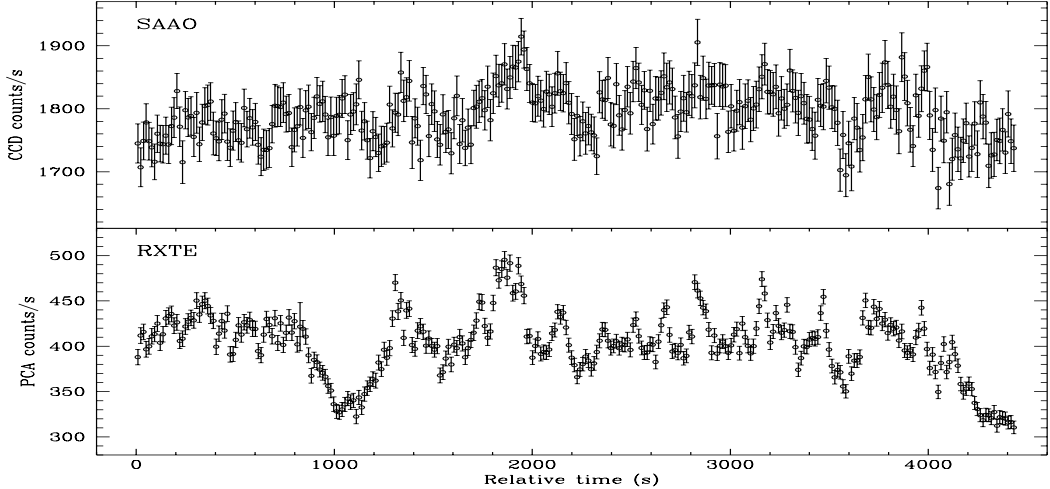}}}
\caption{The optical (top) and X-ray (bottom) light curves of \lmc\ for the run
in which correlated variability is found, taken 1997 December 4.  Both light 
curves have been binned on 16~s.}\label{fig:r303}
\end{figure*}

The X-ray data has a time resolution of 1~s, the optical data 2--10~s. 
Visual examination of the light curves shows that the variability in
both is on timescales of greater than many tens of seconds.  Hence we
found this variability to be displayed most clearly when both X-ray
and optical data were binned into 16~s intervals
(Fig.~\ref{fig:r303}).  However, to quantitatively study the
correlation between the two, and to search for any delays between the
two wavebands, we binned the X-ray light curve onto the corresponding
optical light curve time bins i.e.\ 2, 5 or 10 s (see Table
\ref{tab:saao_log}).  

\section{LMC~X--2 Temporal Analysis}

\lmc\ exhibits clear night-to-night variations of several tenths of a
magnitude, fine structure within each night is also evident
(Fig.~\ref{fig:opt_xr}).  The intervals of optical photometry are too
short to allow detection of the previously quoted orbital periods of
8.15~h and 12.5~d.  Furthermore, a period search on the dataset failed
to reveal any consistent shorter term periodicities in the data.

In order to investigate the correlation between the optical and X-ray
data, to determine whether a delay is present and to quantify this
delay we performed (i) a cross-correlation analysis, and (ii)
modelling of the optical light curve by convolving the X-ray light
curve with a Gaussian transfer function.

\subsection{Cross-correlation}
\label{sect:lmc_cross}

We performed the cross-correlations of each run of optical and X-ray
data by employing a modified version of the Interpolation Correlation
Function, ICF \cite{gask87,hynes98}.  As our optical and X-ray data are
binned onto the same time resolution, and we edited the lengths of the
two time series to be the same, interpolation is not required.  The 
cross-correlation technique that results is effectively a Discrete
Correlation Function, DCF, similar to that of Edelson \& Krolik
(1988).  We also implemented code based on standard cross-correlation
function (CCF) routines which are again similar to a DCF; both the ICF
and CCF methods agree well.  

Concerns have been raised in the literature (see Koen 1994) about the
validity of cross-correlation as a means to finding lags within
wavebands for a source due to the effects of the auto-correlation
function in the individual time-series on the CCFs produced.  This can
lead to the peak in the CCF being shifted from its true value.

\begin{figure*}
\resizebox*{0.5\textwidth}{.25\textheight}{\rotatebox{90}{\includegraphics{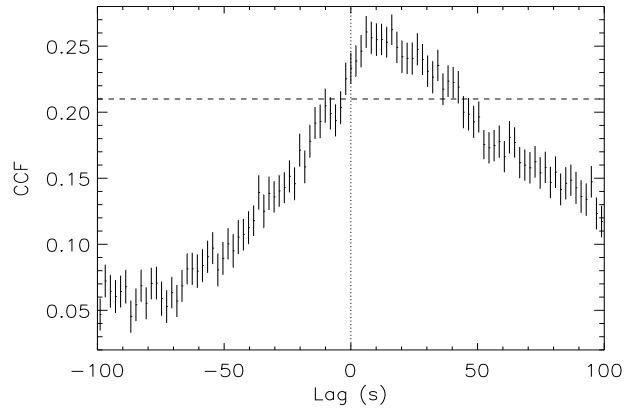}}}
\caption{Cross-correlation function for \lmc.  Sign convention is such 
that positive lags correspond to X-rays leading the optical.  The dashed line 
shows the $3\sigma$ significance level of the CCF from the standard
errors (see Koen 2003).}\label{fig:r303_ccf}
\end{figure*}

For the one simultaneous run (JD 2450787.355 -- 2450787.406) where \lmc\
is seen to be in a bright, active X-ray state (i.e.\ flares are
present and the mean PCA count rate in the 2--10~keV light curve is
400 counts/s) a broad peaked CCF is produced.  For this run the
resulting time bins are of 2.019~s.  The position of the CCF peak
indicates there is a non-zero delay of order $\sim20$~s between the two 
datasets.  The sign convention of the CCF implies that the optical
lags the X-rays (Fig.~\ref{fig:r303_ccf}).  This is expected if the
delay is attributed to the X-rays heating some part of the binary
system which in turn produces optical emission.  The standard
deviation for the CCF for uncorrelated data is given by
$\sigma_{n}=1/(n-2)^{1/2}$, where $n$ is the number of observed points
\cite{gask87}.  However, this is only valid for data with no
autocorrelation.  We therefore use the results from Koen (2003) in
which the standard errors on the CCF are determined by fitting
standard parametric times series models to the data.  In all other runs
the X-rays were in a low state (with mean PCA count rate for the
2--10~keV light curve $<400$~counts/s) and no significant peaks in
the CCFs are found.      

As noted above, although the X-ray light curves have an
intrinsic 1~s time resolution, the data generated by the background 
subtraction models for the X-ray light curve have a time resolution of
16~s.  In order to determine whether or not the background model is 
contributing to the delay in the CCF, non-background subtracted X-ray
data were also used.  We detrended these data to reduce the long-term
effects of the changing background before the cross-correlation
analysis was performed.  We find a broad peak in the CCF with similar
values as previously, indicating that the lag we find is not due to
the background subtraction procedure.  

To calculate a mean value for the range of delays found for \lmc, and
to investigate its significance, we performed Monte Carlo simulations
to create simulated optical and X-ray light curves.  The simulated
data were produced by sampling a Gaussian random number generator with
a mean value equivalent to the average of the entire real dataset, and
a sigma equal to the error bar on each individual real point.
Employing this method we calculated new values for each observed
optical and X-ray data point.  The simulated datasets were
cross-correlated with each other exactly as for the real data.  To
produce good statistics, this was repeated ten thousand times.
Statistics were computed on the spread of values obtained giving a
mean of 14.2~s and a standard deviation of 8.7~s.  This value
represents the typical time lag for these data, however, a typical lag
of zero is not excluded at the level of twice the standard deviation.
A zero delay is {\em not} ruled out as there {\em are} times when
there is {\em no} lag, and we expect it to be modulated on the orbital
period.  Hence the quoted error ($\pm8.7$~s) is {\em not} the true
error on the mean delay, but more an indication of the range over
which the lags are seen.  We note that this method leads to there
being no correlation between each set of simulated optical and X-ray
data.  Therefore we have a measure of the spread of correlations that can
occur due to chance when there is no real correlation between the two
datasets.  A much more detailed study of the statistical properties of 
cross-correlating limited datasets with variable binning is presented
by Koen (2003), in an analysis which uses some of these data of \lmc\
as an example.  As for the standard analysis we use here, Koen (2003) 
concludes that there is strong evidence for variable X-ray/optical
lags, but is able to demonstrate the level of statistical significance
much more clearly. 

\begin{figure*}
\resizebox*{0.85\textwidth}{0.45\textwidth}{\rotatebox{90}{\includegraphics{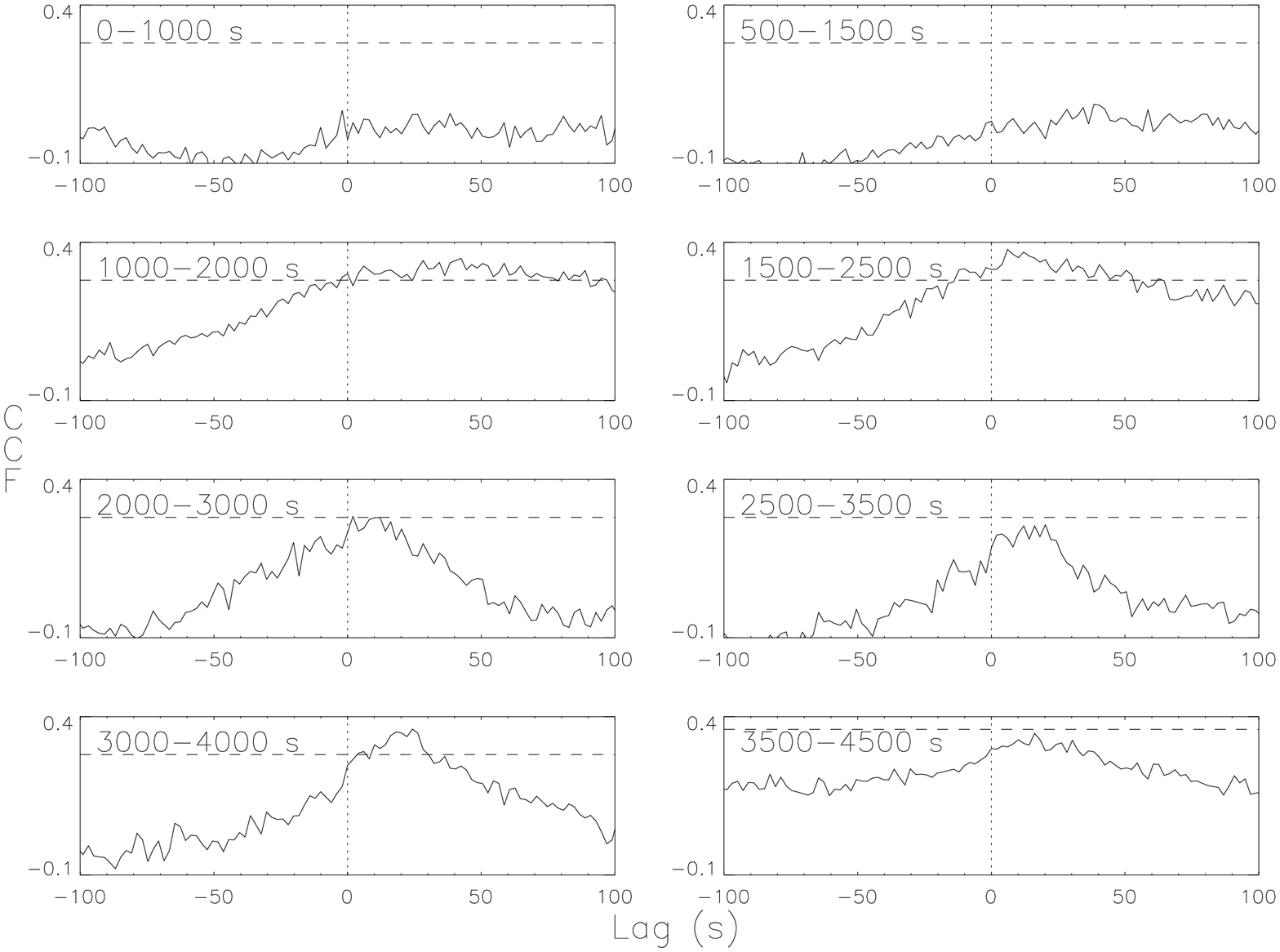}}}
\caption{Cross-correlation functions for 1000 s sections of optical 
and X-ray data of \lmc.  Sign convention is such that positive lags correspond 
to X-rays leading the optical.  The dashed lines show the $2\sigma$
significance level of the CCF from the standard errors (see Koen
2003).}\label{fig:sect}
\end{figure*}

To study the effect of the flaring behaviour in the 2--10~keV X-ray
light curve we cross-correlated 1000~s sections of the optical and
X-ray data as above.  The CCFs produced (Fig.~\ref{fig:sect})
demonstrate well that correlated variability is only seen when there
is high X-ray activity.  The first three panels (0--2000~s) in
Fig.~\ref{fig:sect} show the effect that the large dip in the X-ray
light curve has on the correlated variability (which is to produce
the longer lags).  The other panels (1500--4500~s) which have no
contribution from the dip show much narrower CCFs. 

\begin{figure*}
\resizebox*{.48\textwidth}{.35\textheight}{\rotatebox{90}{\includegraphics{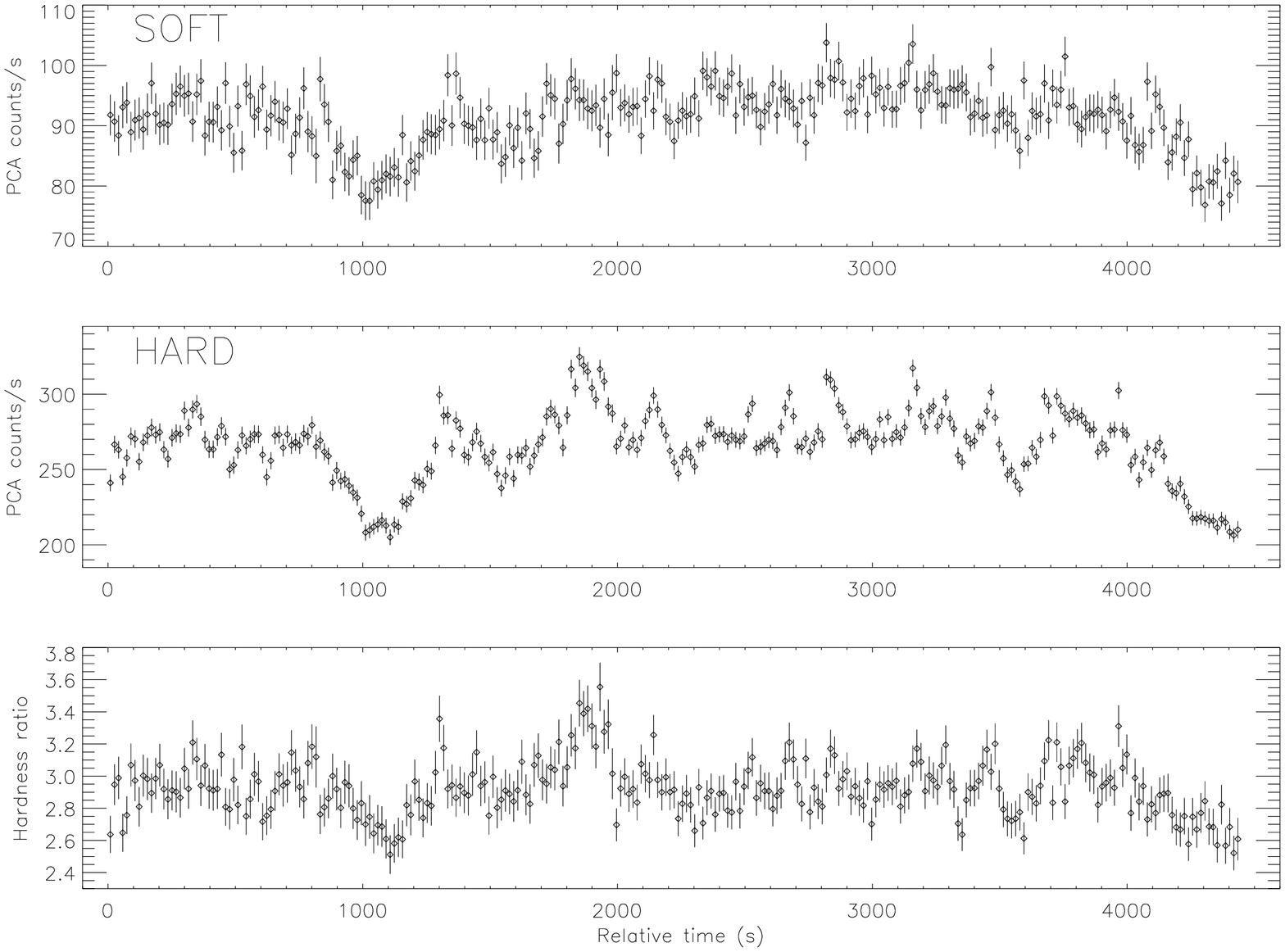}}}
\resizebox*{.48\textwidth}{.35\textheight}{\rotatebox{90}{\includegraphics{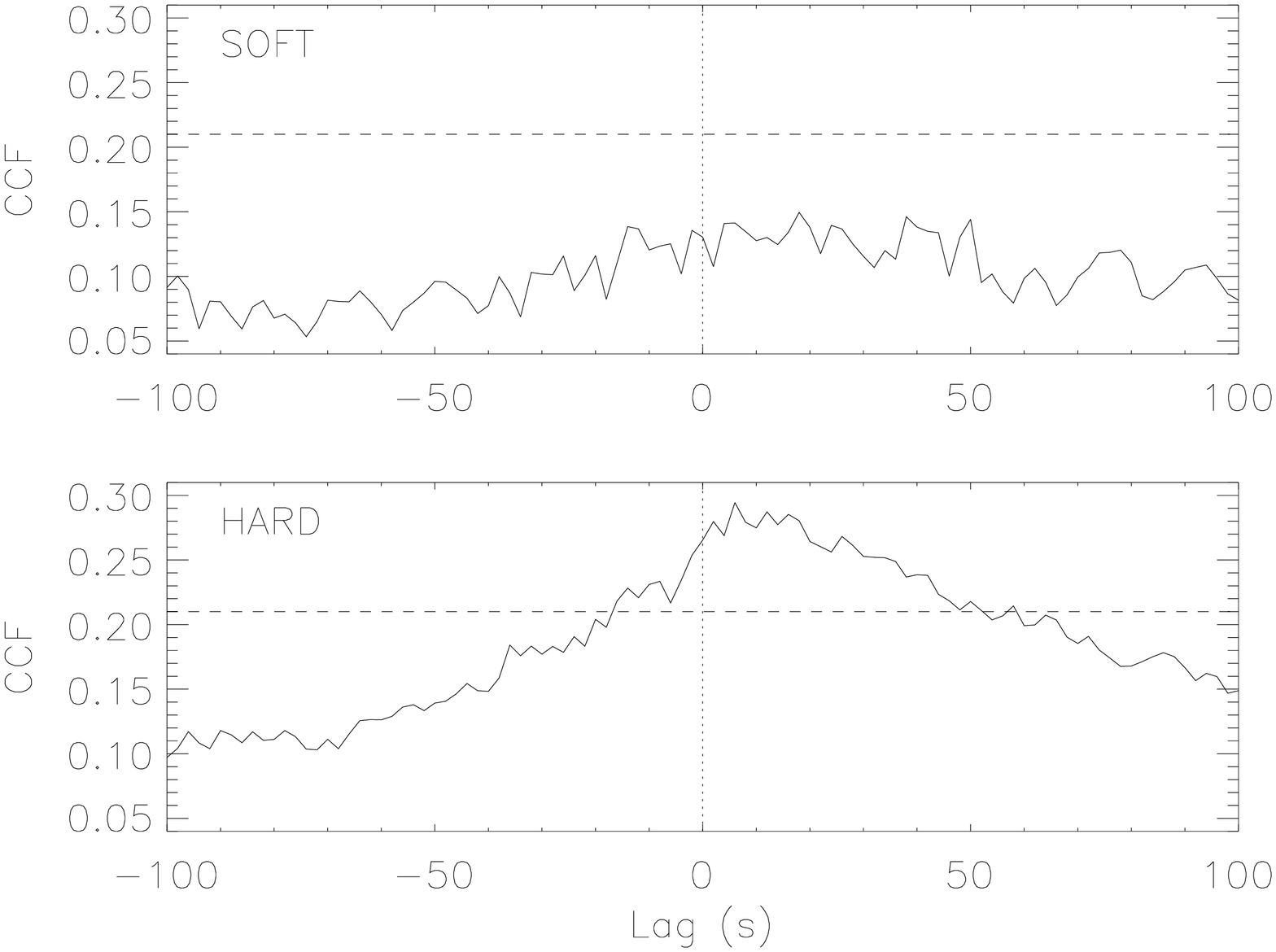}}}
\caption{Left panels, soft (2--4~keV) (first) and hard (4--10~keV)
(second) X-ray light curves of \lmc, hardness ratio (third), defined as 
(hard)/(soft).  Right panels, the CCFs for the soft (first) and hard
light curves (second).  Sign convention is such that positive lags
correspond to X-rays leading the optical.  The dashed lines show the 
$3\sigma$ significance level of the CCF from the standard errors (see Koen
2003).}\label{fig:hard_soft}
\end{figure*}

We also investigated the source of the flaring activity observed
in the X-ray light curve.  We extracted data in the soft (2--4~keV)
and hard (4--10~keV) energy ranges (see Section 2).  The light curves
were cross-correlated with the optical data as before.  The hard and
soft X-ray light curves are shown in Fig.~\ref{fig:hard_soft} (left).
The light curves clearly demonstrate that \lmc\ is most variable at
hard energies with little activity in the soft energies, as has been
noted by Bonnet-Bidaud et al.\ (1989).  The CCFs for both light curves
are shown in Fig.~\ref{fig:hard_soft} (right).  We note that there are
less counts in the soft X-ray light curve than the hard X-ray light
curve which could be responsible for the lack of a significant peak in
the CCF for the soft band.  However, we find that in the hard light curve
the feature at $\sim1800$~s has a $\sim16$\% excess compared to the
mean, while in the soft light curve it is an excess of only $\sim5$\%.
This indicates that the features which lead to the CCF peak in the
hard band are not present in the soft energies.  The large dip is
present in both the hard and soft light curves.    

The measurement of the cross-correlation function provides a
characteristic delay which does not depend on particular model
fitting.  To characterize the distribution of time delays present
between the optical and X-ray light curves we modelled the data with a
transfer function. 

\subsection{Transfer Function}
\label{sect:lmc_gauss}

In order to model the time lag between the optical and X-ray data of
\lmc\ we predict the optical light curve by convolving the observed
X-ray light curve with a Gaussian transfer function.  We use
$\chi^{2}$ fitting to obtain the best-fit to the optical light curve
(see Hynes et al. 1998; Kong et al. 2000).  The Gaussian transfer
function is given by
\begin{equation}
\label{eq:gauss}
\psi(\tau) = \frac{\Psi}{\sqrt{2\pi\Delta\tau}} e^{-\frac{1}{2}\left
(\frac{\tau-\tau_{0}}{\Delta\tau}\right )^2} ,
\end{equation}
where $\tau_{0}$ is the mean time delay, and $\Delta\tau$ is the
dispersion or root-mean-square time delay, which is a measure of the
width of the Gaussian, and is equivalent to the degree of 'smearing'.
The strength of the response is given by $\Psi$. 

\begin{table}
{\caption{\label{tab:gauss} Summary of results from convolution of a
Gaussian transfer function to the three different X-ray energy band
light curves of \lmc.}}
\begin{center}
\begin{tabular}{l|c|c}
\hline
{} & {2--10~keV} & {4--10~keV} \\
\hline 
$\tau_{0}$ (s) & $18.6_{-6.6}^{+7.4}$ & $14.7_{-5.7}^{+7.3}$ \\
$\Delta\tau$ (s) & $10.2_{-5.7}^{+5.8}$ & $9.0_{-4.5}^{+7.0}$ \\
$\Psi (10^{-3})$ & 10.8 & 13.4 \\
$\chi^{2}_{\nu}$ & 0.71 & 0.69 \\
\hline
\end{tabular}
\end{center}
\end{table}

\begin{figure*}
\resizebox{0.7\textwidth}{0.4\textheight}{\rotatebox{90}{\includegraphics{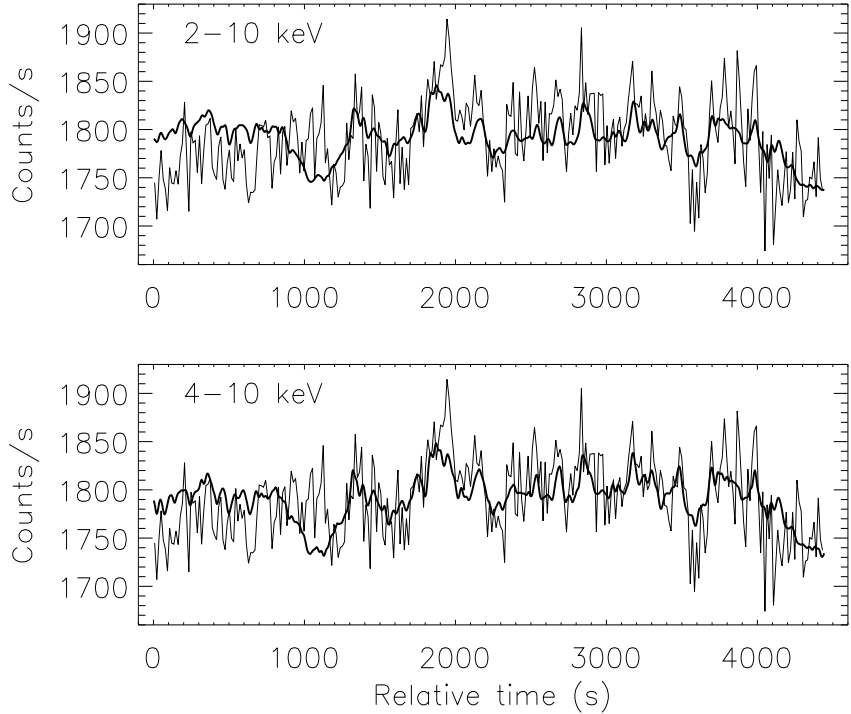}}}
\caption{Best-fitting predicted light curves (thick line) using a
Gaussian transfer function on the two X-ray bands of \lmc, 2--10~keV
(top) and 4--10~keV (bottom).  The resulting curves are superimposed
on the optical light curve which has been binned on 16~s for clarity.}\label{fig:gauss}
\end{figure*}

We performed a series of convolutions of the transfer function with
the X-ray light curve in the 2--10~keV and 4--10~keV energy bands,
varying both $\tau_{0}$ and $\Delta\tau$ independently.  
Table~\ref{tab:gauss} summarizes the results of fitting Gaussian transfer
functions to the two X-ray light curves.  Fig.~\ref{fig:gauss} shows
the best-fitting predicted light curves from the convolutions
superimposed on the optical light curve of \lmc.  We did not perform
this analysis with the soft (2--4~keV) X-ray data due to the lack of
variability in the light curve (see Section~\ref{sect:lmc_cross}).

The values for $\chi_{\nu}^{2}$ are good for both bands.  The
principal features of the optical light curve are reproduced well in
the predicted light curves from the 2--10~keV and 4--10~keV energies.

The mean delay and dispersion between the optical and 2--10~keV X-ray
data are 18.6 and 10.2~s respectively, for the 4--10~keV data the mean
delay and dispersion are 14.7 and 9.0~s.  The errors on the mean delay
values indicate that a non-zero lag is present at the $2.8\sigma$ and 
$2.6\sigma$ levels for the 2--10~keV and 4--10~keV bands.  The
strength of the response is an indication that a greater proportion of
the reprocessing is driven by higher energies.

\section{Comparison with Sco~X--1}

\begin{figure*}
\resizebox*{.46\textwidth}{.53\textheight}{\rotatebox{0}{\includegraphics{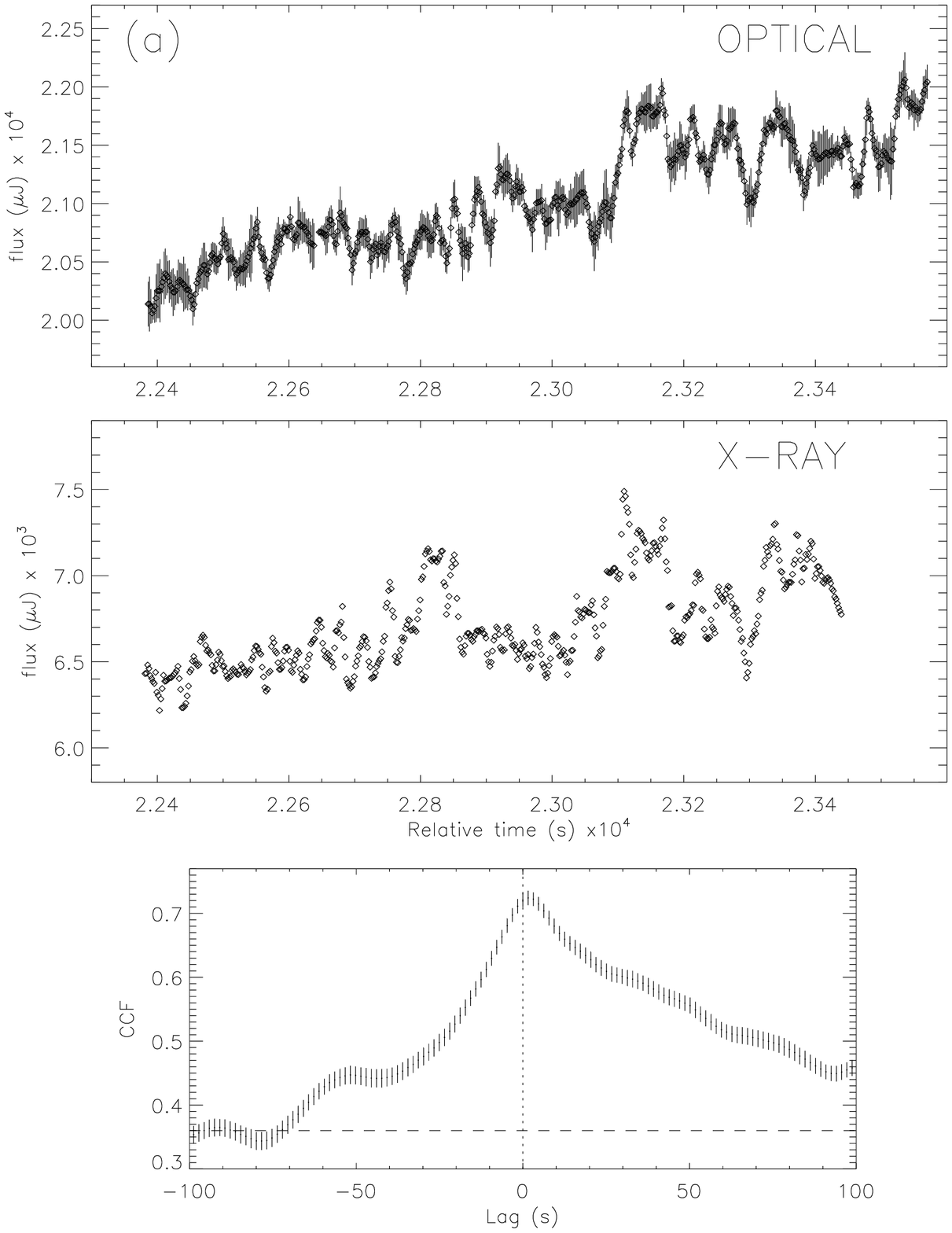}}}
\resizebox*{.46\textwidth}{.53\textheight}{\rotatebox{0}{\includegraphics{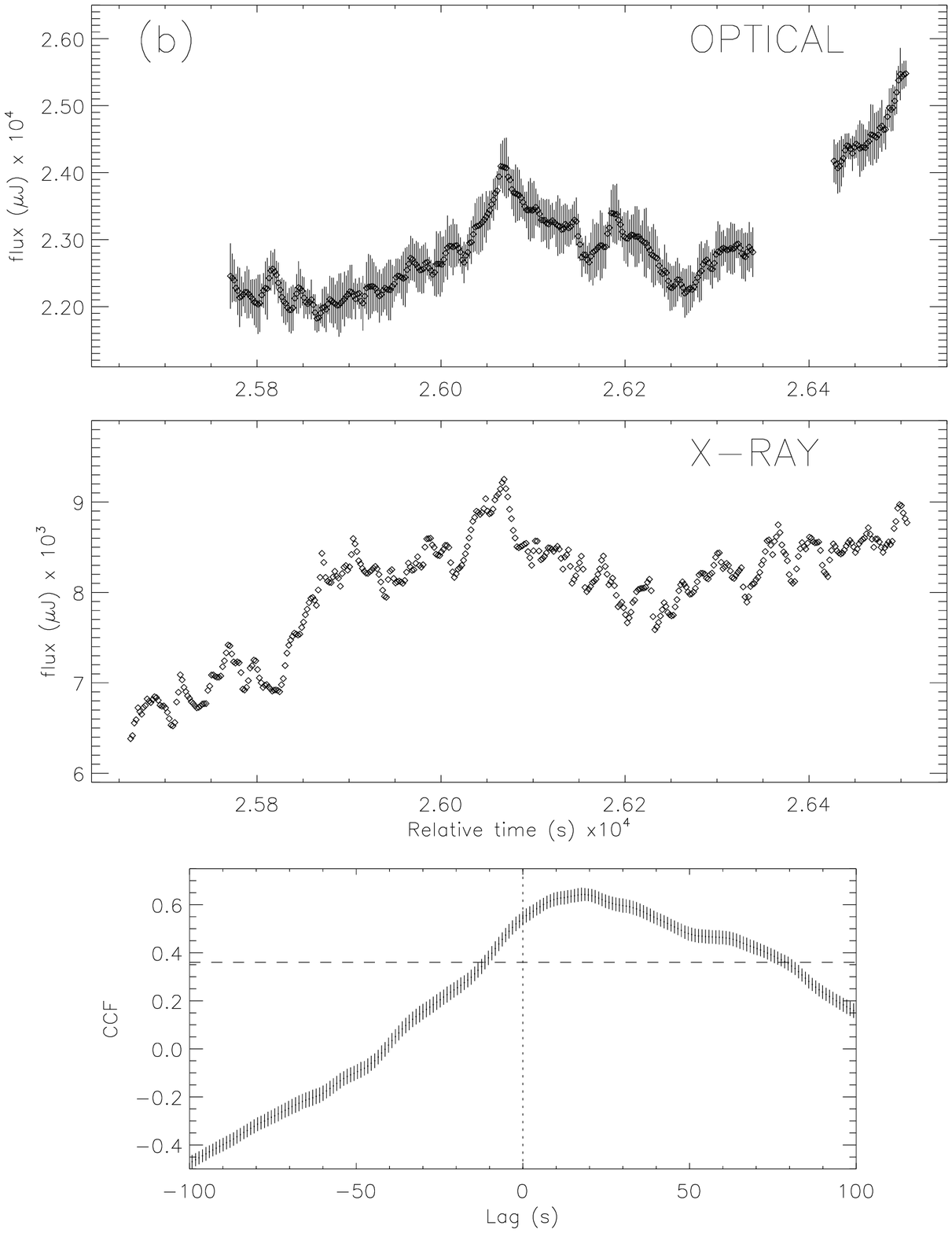}}}
\resizebox*{.5\textwidth}{.38\textheight}{\rotatebox{90}{\includegraphics{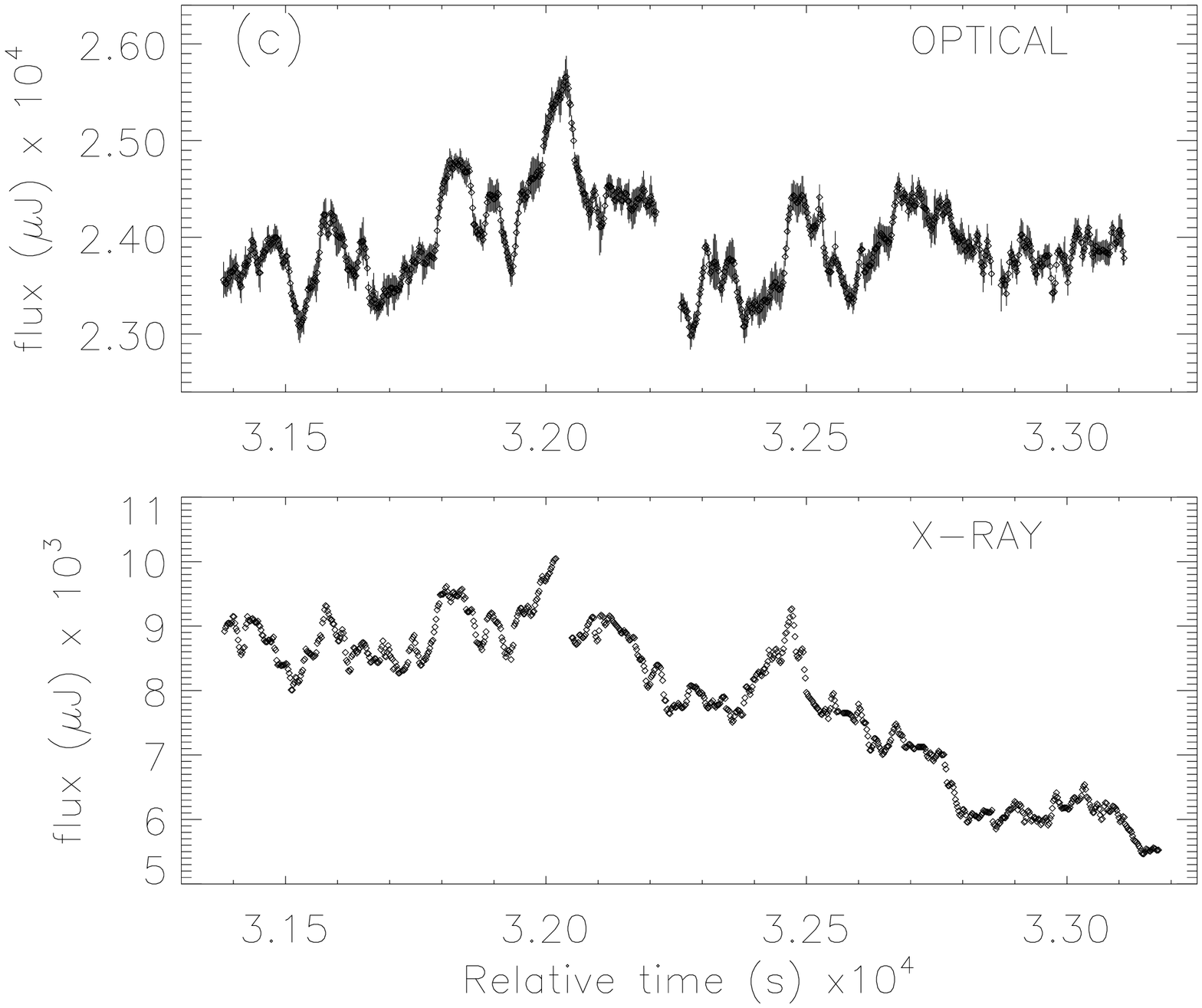}}}
\caption{(a), (b), (c) Simultaneous optical and {\em SAS-3} X-ray
active state data of \sco\ taken on 1977 March 15 (Ilovaisky et al.\
1980).  The lower panels in (a) and (b) show the CCFs for these two
datasets.  All light curves are binned on 2~s resolution.  The dashed
lines in the CCFs in (a) and (b) show the $3\sigma$ significance level
of the CCF from the standard errors (see Koen 2003).}\label{fig:ilov}
\end{figure*}

It is instructive to compare \lmc's behaviour with that of the brightest
galactic LMXB, \sco.  Ilovaisky et al.\ (1980; hereafter I80) studied
the correlation of variability in simultaneous optical and X-ray
observations of \sco\ using {\em SAS-3}, {\em Copernicus} and
conventional photoelectric photometry.  As we have found with \lmc,
I80 only detected correlated variability occurring when the source was
in a bright, active X-ray state (Fig.~\ref{fig:ilov}), with the
correlation decreasing as \sco\ became less active.  More
interestingly, they also found that the peak in their
cross-correlation curve gave a delay which was consistent with the
optical being delayed with respect to the X-rays.  They attributed
this delay to reprocessing of the X-rays in the binary system.  From
their observations during a period of high activity they suggested
that the optical features were delayed by $\sim30$~s compared with
those in the X-rays.  From their cross-correlation analysis for those
data when \sco\ was in an active state they found a range of values
for the delay from a few seconds to a few tens of seconds.  The peak
value for the cross-correlation function (CCF) at these delays was not
much greater than that for zero lag and so they concluded that the
delays were not very significant.    

Petro et al.\ (1981; hereafter P81) also found correlated variability
in \sco.  Their analysis suggested an optical smearing timescale
('filter' or 'processing' time) of $\sim20$~s, describing the optical
flares as 'filtered' versions of the X-ray flares.  They concluded
that the delay could not be due to reprocessing on the secondary as
this required a timescale of only $\sim10$~s, and therefore could not
explain the presence of the long smearing timescale for the system.
From the rise times of the correlated optical and X-ray flares they
also concluded that the reprocessing site that produced the optical
flares must be in a different region from the site of X-ray production
(c.f.\ Pedersen et al.\ 1982).   

Following our analysis above of \lmc, we decided to investigate such
previous X-ray/optical observations of LMXB's in greater detail.  

\begin{figure*}
\resizebox*{.4\textwidth}{.5\textheight}{\rotatebox{0}{\includegraphics{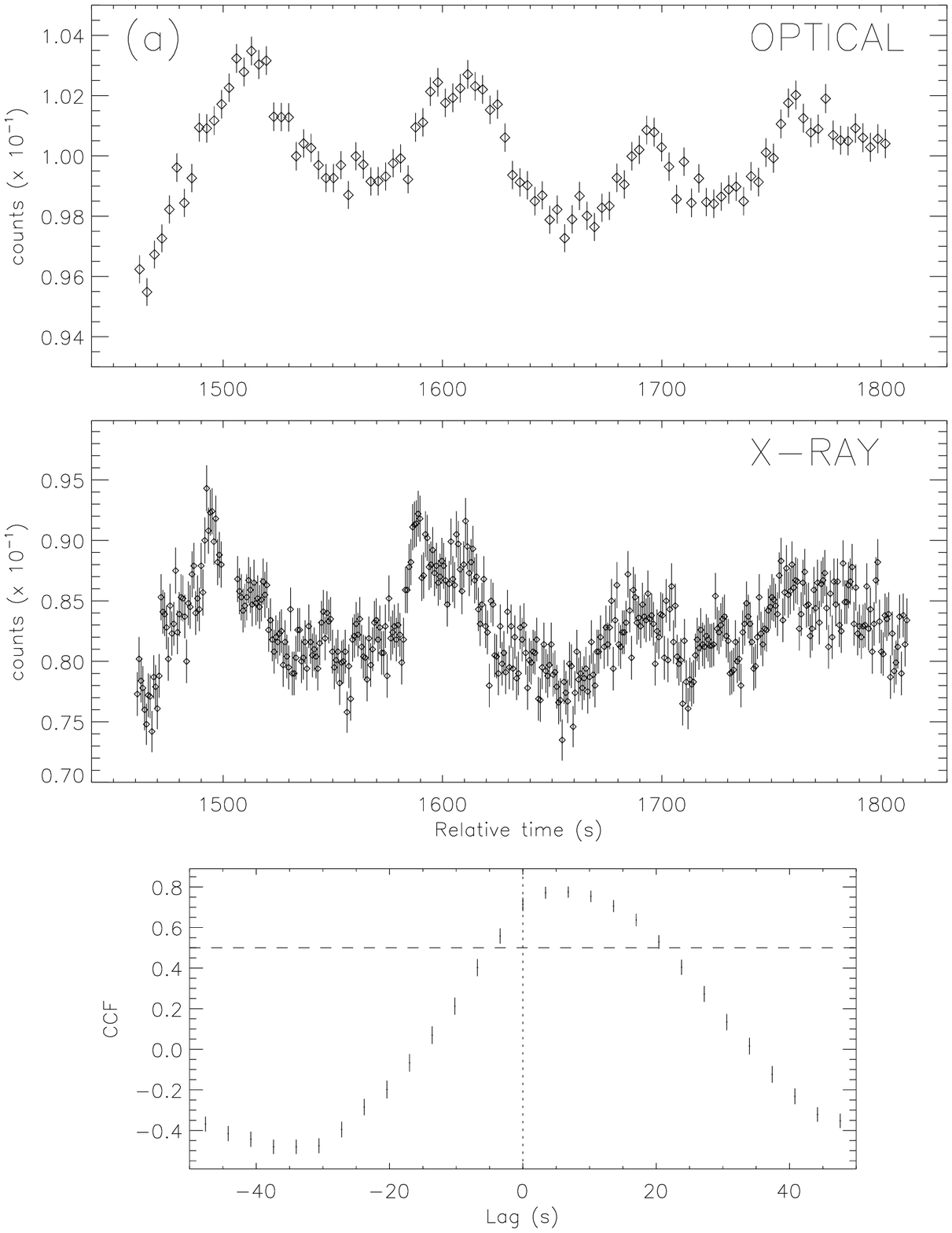}}}
\resizebox*{.4\textwidth}{.5\textheight}{\rotatebox{0}{\includegraphics{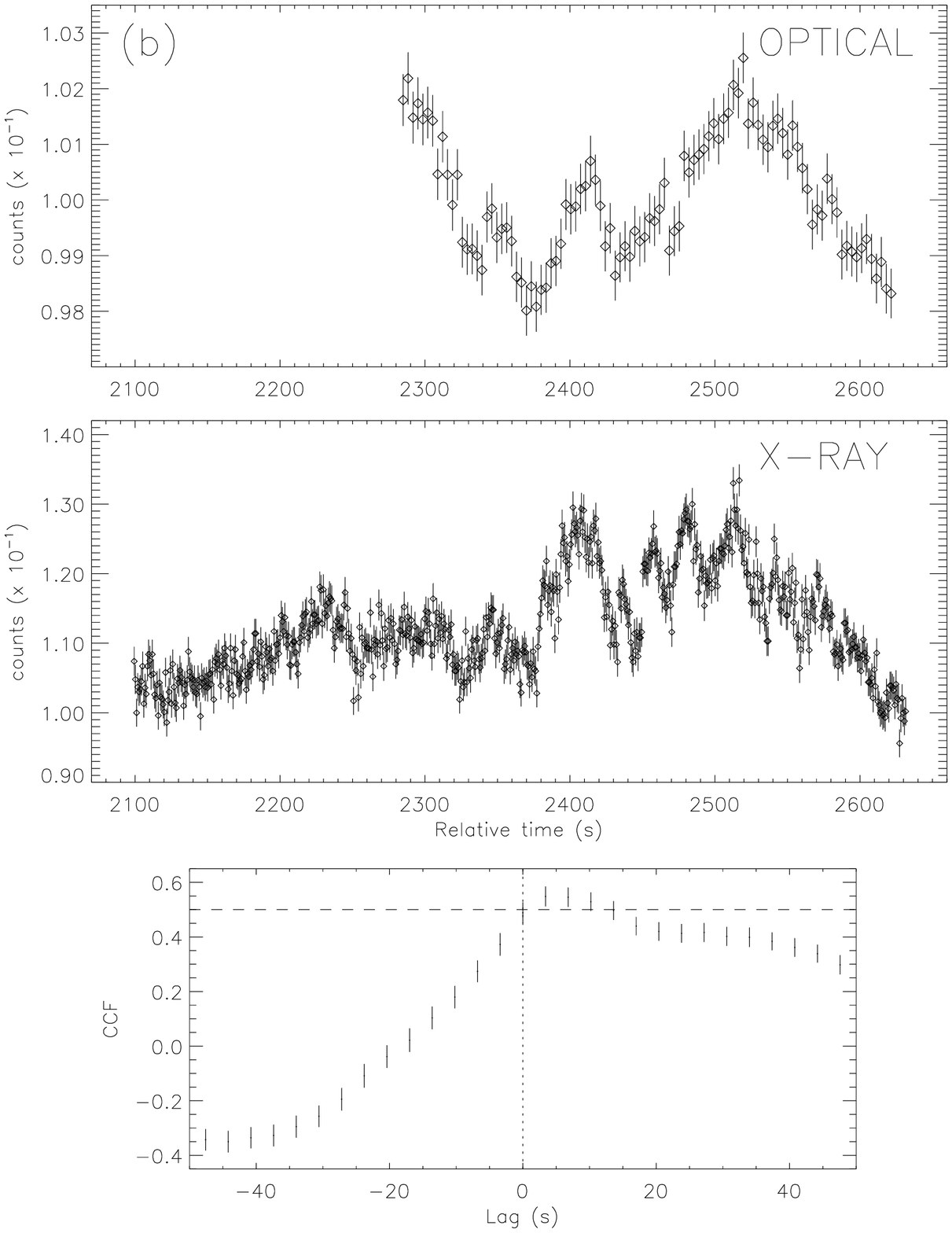}}}
\caption{(a), (b) Simultaneous optical and X-ray active state data of
\sco\ taken on 1979 March 8 (Petro et al.\ 1981).  The lower panels
show the CCFs for these two datasets.  The dashed lines in the CCFs
show the $2\sigma$ significance level of the CCF from the standard
errors (see Koen 2003).}\label{fig:petro}
\end{figure*}
   
\subsection{Cross-correlation}

We digitized the active state optical and X-ray data of \sco\ from
I80, converting the data to $\mu$Jy.  We then performed
cross-correlations of the simultaneous data as have been done for
\lmc\ (Fig.~\ref{fig:ilov}).  We find that the delays for the first
two sets of data are $\sim1.7\pm0.9$~s and $17.6\pm5.4$~s.  The
significance of both CCF peaks is $>4\sigma$.  The results indicate
that a lag of zero is not ruled out at the $2\sigma$ level for the
first set of data, for the second set there is $>3\sigma$ confidence
of a non-zero lag.  For the third set of data no significant delay is
found, but note the decline in the X-rays compared to the constant
optical.  The lack of a delay for the second half of the data
demonstrates again that the X-rays must be in an active state (i.e.\
with flaring) to show correlated variability with the optical data.       
  
We cross-correlated the two simultaneous active state datasets from
P81, finding delays of $5.9\pm2.5$~s and $5.7\pm2.8$~s
(Fig.~\ref{fig:petro}).  There is evidence for a non-zero lag in both
datasets at the $>2\sigma$ level.  

As in \lmc, \sco\ exhibits an increase in correlation between the
optical and the X-rays as the source becomes more active.  I80
suggested that this could be due to either one or two regions of
optical emission.  In the case of one optical emitting zone, they
suggested that the optical reprocessing region is illuminated by the
X-ray source at all times, but 'something' is required to damp out the
X-ray variability between the compact object and the observer when the
source is in a non-active state, or perhaps the optical ``region''
cannot then see the X-rays.  For the case of two emitting regions,
they suggest that the uncorrelated emission ($\sim25$\% less energetic
in the X-rays than during correlated emission) could be due to optical
emission from the disc itself, while the correlated emission could be
due to X-ray heating of different parts of the system and would be
directly related to the X-ray emission.  

The lags determined from the cross-correlation analysis again
represent the characteristic delay present between the optical and
X-ray light curves.  In order to model the time delay we have also
convolved the X-ray light curves of \sco\ from I80 and P81 with a
Gaussian transfer function, as has been done for \lmc\ (see
Section~\ref{sect:lmc_gauss}). 

\begin{figure*}
\resizebox{0.7\textwidth}{0.4\textheight}{\rotatebox{90}{\includegraphics{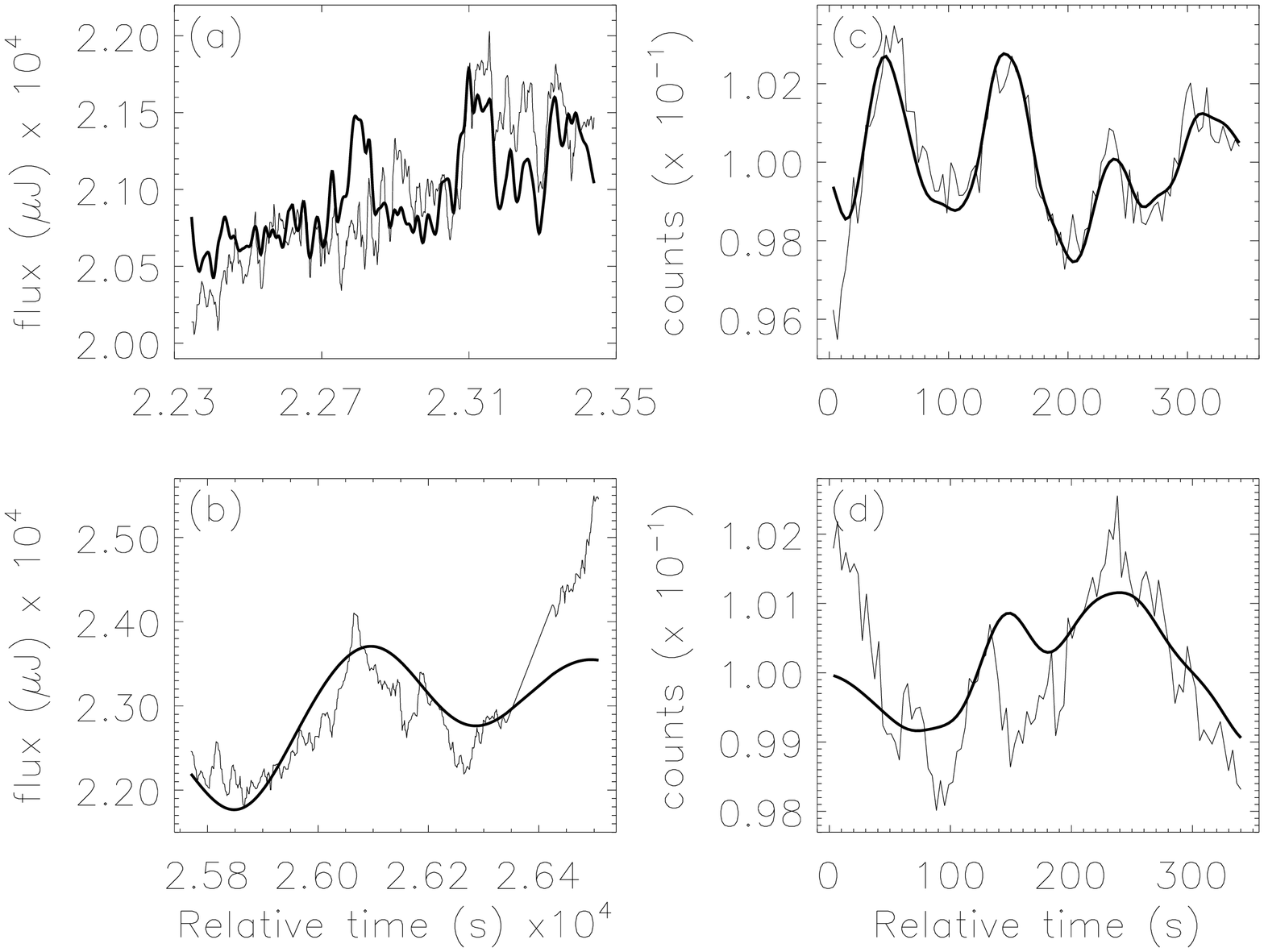}}}
\caption{Best-fitting predicted light curves (thick line) using a
Gaussian transfer function on the X-ray light curves of \sco\ from
Ilovaisky et al.\ (1980) (a), (b) and from Petro et al.\ (1981) (c),
(d).  The resulting curves are superimposed on the optical light curves.}\label{fig:gauss2}
\end{figure*}

\subsection{Transfer Function}

A series of convolutions of the transfer function with each of the
X-ray light curves were performed on the first two datasets from
I80 (Fig~\ref{fig:ilov} (a) and (b)), and both sets of data from P81 
(Fig~\ref{fig:petro} (a) and (b)).  Again both $\tau_{0}$ and
$\Delta\tau$ were varied independently.  The results of fitting the
Gaussian transfer functions to the \sco\ X-ray data are summarized in
Table~\ref{tab:gauss2}.  The values of $\chi_{\nu}^{2}$ indicate that
the fits are not good.  The best-fitting predicted light curves from
the convolutions superimposed on the optical light curves are shown in
Fig~\ref{fig:gauss2}. 

\begin{table}
{\caption{\label{tab:gauss2} Summary of results from convolution of a
Gaussian transfer function to the X-ray light curves of \sco.}}
\begin{center}
\begin{tabular}{l|c|c|c|c}
\hline
{} & {Ilovaisky} & {Ilovaisky} & {Petro} & {Petro} \\
{} & {(a)} & {(b)} & {(a)} & {(b)} \\
\hline 
$\tau_{0}$ (s) & $1.6_{-0.5}^{+0.6}$ & $42.0_{-3.0}^{+2.5}$ & $8.0\pm0.8$ & $21.9_{-4.9}^{+5.1}$ \\
$\Delta\tau$ (s) & $4.9_{-0.7}^{+0.8}$ & $82.0_{-4.0}^{+5.0}$ & $8.6\pm1.3$ & $16.8_{-3.8}^{+4.2}$ \\
$\Psi (10^{-3})$ & 5.27 & 5.30 & 74.7 & 13.8 \\
$\chi^{2}_{\nu}$ & 4.00 & 4.05 & 2.98 & 3.69 \\
\hline
\end{tabular}
\end{center}
\end{table}

The predicted light curves reproduce the optical data marginally for
the first set of I80 data and the first set of P81 data
(Fig~\ref{fig:gauss2}, (a) and (c)).  For the second sets of data from
I80 and P81 (Fig~\ref{fig:gauss2}, (b) and (d)) only the overall shape
of the optical light curves are reproduced, and not the small scale
features.  The results for the first datasets of I80 and P81 suggest
non-zero lags of 1.6 and 8.0~s respectively, and are present with
$3.2\sigma$ and $10\sigma$ confidence, respectively.

\section{Discussion}

We find evidence that the optical emission of \lmc\ is delayed with
respect to the X-ray emission.  The presence of a non-zero delay has
$>2.6\sigma$ confidence.  If the lag is greater than zero, reprocessing
of the X-rays in material at some distance from the X-ray source could
be responsible.  Potential candidates as reprocessing regions include
the outer areas of the accretion disc and the heated face of the
secondary star.  However, to determine where the reprocessing is
occurring requires a detailed knowledge of the binary system
parameters.  These same {\em RXTE} observations have provided a
possible new confirmation of the 8.2~h orbital period in the X-ray
data \cite{smale00}, but the nature of the compact object remains
obscure.  Assuming that 8.2~h is indeed the orbital period (see also
Alcock et al.\ 2000), then the compact object could either be :   

(a) a neutron star, assumed mass 1.4\msun\ in which case the tidal
disc radius will be $\sim1$\rsun\ (equiv. 2--3~s light travel time)
and the secondary at a distance $\sim2.5$\rsun\ ($\sim6$~s); or (b) a
10\msun\ black hole with disc radius $\sim2$\rsun\ ($\sim4$--5~s) and 
separation of the secondary $\sim5$\rsun\ ($\sim11$~s).  [Note,
however, that if the period were to be 12.5~d, then these delays would
be (a) 25--28~s and $>60$~s (b) 47--54~s and $>117$~s].  

The light travel times arise from the time of flight differences for
emission that is observed directly and emission which is reprocessed
and re-emitted before travelling to the observer.  The maximum delay
is twice the binary separation, plus any time due to
reprocessing/diffusion.  Our results for the optical and
2--10~keV X-ray data of \lmc\ give a mean lag of
$18.6^{+7.4}_{-6.6}$~s and a distribution of $10.2^{+5.8}_{-5.7}$~s.
The $2\sigma$ lower limit for the lag is 5.4~s.  Employing the period and
ephemeris given in Smale \& Kuulkers (2000), the phase at the
beginning of the correlated optical and X-ray observations of \lmc\ is
0.4, where $\phi=0$ corresponds to the time of minimum light.  Lags of
greater than 6~s would be expected for this phase if the secondary
star is the reprocessing site, thus the $2\sigma$ lower limit for the
lag may be reconciled with disc reprocessing.  

The light travel times for the disc and the secondary star in \sco,
given its known orbital period of 18.9~h \cite{gott75,lasala85}, are
4--5 and 10~s respectively.  Note that the secondary has never been 
detected in \sco, however QPO have been observed \cite{middle85} and
the delay values given above assume a neutron star compact object.
Considering the first sets of \sco\ data from I80 and P81, our
$2\sigma$ lower limits for the lags of 0.6 and 6.4~s from the
convolution of the X-ray data with a Gaussian transfer function
suggest reprocessing in the disc.  However, the goodness of fits are
poor, and the transfer function does not work well for the other two
sets of \sco\ data.

In \lmc\ and \sco\ the delays found from Gaussian modelling and
the cross-correlation analysis are in good agreement.  It is therefore
likely that the non-zero delays determined for the second sets of data
from I80 and P81 with the cross-correlation technique are close to the
true values.  In both cases the lower limit for the lags are
consistent with the disc being the site of reprocessing.

Other authors have found delays in optical/X-ray bursts for X-ray
binaries in which the optical emission is suggested to be the result
of X-ray reprocessing.  The optical burst is described as a delayed
and smeared version of the X-ray burst, the time delays found being
consistent with the reprocessing occurring in the accretion disc
(Matsuoka et al.\ 1984 and references therein; Kong et al.\ 2000).
The time delays found for the bursts are much shorter than those for
\lmc.  However, the X-ray spectra for the bursts are much softer than
for \lmc, indicating that the delays found for the bursts are due to
processes closer to the surface of the disc than we are seeing.  We
find evidence for time lags when \lmc\ displays hard flaring behaviour
which suggests that the optical is responding to X-rays which have
penetrated to a deeper level in the disc. 

However, for \lmc\ and \sco\ we are considering the $2\sigma$ lower
limit for the lags, which implies that our delays could be much
longer.  In the case of \lmc, even the lower limit for the delay is
longer than the light travel time of the disc, while shorter than the
distance to the secondary.  Note that this result is from only one
correlated optical and X-ray run. 

It is obvious that more data are needed to confirm this result.  It
would also be instructive to use the technique of O'Brien et
al. (2002) in which the time delay transfer functions are modelled by
simulating the distribution of the reprocessing regions, employing
geometrical and binary parameters.  In order to explain the longer
than disc crossing times however, more sophisticated radiative
transfer models are likely needed.  We require some mechanism that can
add a finite time to the light travel time to account for the longer
delays that may be present.  One possible method is to take into
account the time component that could be present due to diffusion.
Thus, the delay could be interpreted as the light travel time with an
added component due to diffusion within the absorption/re-emission
region.  Calculations to determine the order of the diffusion
timescale using a simplistic, however unrealistic, single temperature
atmosphere are presented in Appendix~A. 

Our observations of \lmc\ show the first correlated optical and X-ray
variability for the source.  The temporal analysis provides evidence
that the X-rays lead the optical, which implies reprocessing is
occurring.  We also find that the optical light curves of \sco\ are
delayed with respect to the X-ray light curves from analysis of
archival data.  Further observations are needed when \lmc\ and \sco\
are in a bright, active X-ray state to investigate the source of the
lags.  This could be combined with additional optical observations to
study the colours and spectra of the two sources at that time. 

\section{Acknowledgements}

We thank Philipp Podsiadlowski and Keith Horne for useful discussions
on the reprocessing physics.  We also thank Chris Koen, Rob Hynes
and Albert Kong for useful comments on the temporal analysis methods.

\appendix
\section{Diffusion Timescale Calculations}

Using a simple model for the secondary star or the 'atmosphere' of the
disc, we can estimate the duration of the diffusion timescale, the time
it takes for the X-ray photons to be absorbed and the energy deposited
there to be re-emitted in the optical.  This timescale is heavily
dependent on the density and opacity of the star/disc. 

Assume the X-rays penetrate to a depth $d$ in the photosphere
where $\kappa_{x}$ is X-ray opacity ($\kappa_{x}=0.2(1+X)\sim 0.34$
cm$^{2}$g$^{-1}$) and $\rho$ is density, then 
\begin{equation}
\label{eq:d}
d \sim \frac{1}{\kappa_{x}\rho} 
\end{equation}
The number of scatterings is then $\sim (d\kappa_{\circ}\rho)^{2}$
where $\kappa_{\circ}$ is the optical opacity, and the time between each 
scattering is $\sim 1/(\kappa_{\circ}\rho c)$.
Combining these with Eq.~\ref{eq:d} we can define the diffusion timescale as
\begin{equation}
\label{eq:tdiff_sim}
t_{diff} = \left (\frac{\kappa_{\circ}}{\kappa_{x}}\right )^2 \cdot \frac{1}{\kappa_{\circ}\rho
c}
\end{equation}
From the equations for the optical depth and pressure of the
photosphere, assuming that $\kappa_{\circ}$ is constant in depth $d$,
and that the gravity $g$ remains roughly constant over $d$, then we
can derive an expression for $\kappa_{\circ}$ (assuming an ideal gas) 
\begin{equation}
\label{eq:g}
\frac{\kappa_{\circ}}{g} = \frac{\frac{2}{3}}{\frac{k}{\mu_H m_H}\rho T_{eff}}
\end{equation}
Replacing $g$ in Eq.~\ref{eq:g} with $GM/R^{2}$ and rearranging to get 
an expression for $\kappa_{\circ}\rho$ we can substitute this into 
Eq.~\ref{eq:tdiff_sim} to give
\begin{equation}
\label{eq:tdiff_all}
t_{diff} = \left (\frac{\kappa_{\circ}}{\kappa_{x}}\right )^2 \cdot 
\left [\frac{\frac{3}{2}\cdot \frac{kT_{eff}}{\mu_H m_H}}{\frac{GM}{R}} \cdot 
\frac{R}{c}\right ]
\end{equation} 
By calculating the irradiating flux ($F_{x}$) we can derive the irradiation 
temperature $T_{x}$, given by
\begin{equation}
\label{eq:tirrad}
T_x = \left (\frac{fF_x}{2\sigma}\right )^{1/4} 
\end{equation}
where $f$ is 5--10\% and $\sigma$ is the Stefan-Boltzmann constant.
Taking $T_{eff}$ in Eq.~\ref{eq:tdiff_all} to be the irradiation
temperature, and calculating values for the radius and mass of the
secondary in \lmc, we find a value of $\sim 6\times 10^{-3}$~s for the 
expression in square brackets in Eq.~\ref{eq:tdiff_all}.  Using the
quoted values for the gravity and density of a K star \cite{allen73} a
diffusion time of greater than 20000~s is found (Table
\ref{tab:delay}).  As we are trying to calculate the diffusion time
for the heated secondary we should use the gravity of a K star, but
the density of a B star (e.g.\ the 5.2~h LMXB X\thinspace1822-371 has
a low mass secondary but the heated face shows a $\sim$~B star
spectrum, Harlaftis, Charles \& Horne 1997).  For illustrative
purposes we use a canonical B star density (Allen 1973) to give an
estimate for the diffusion time, resulting in $t_{diff} \sim 1$~s.
Table \ref{tab:delay} illustrates how the optical opacity is heavily
dependent on the density of the star, and in turn how this greatly
effects the estimated value for the diffusion time that results.
Detailed modelling is required to calculate the diffusion time for the
disc.  These calculations could assume a spectral-type of $\sim$~A0
for the disc, and would need to take into account the angle of
incidence of the incoming radiation and the gravity and density of the
disc.    

\begin{table}
{\caption{\label{tab:delay}Diffusion times for different spectral types.}}
\begin{center}
\begin{tabular}{c|c|r|r}
\hline
{Spectral} & {log $N$} & {$\kappa_{\circ}$} & {$t_{diff}$}\\ 
{Type} & {(cm$^{-3}$)} & {} & (s) \\
\hline 
B0  &  15.0  &  4.1  &  1 \\
A0  &  15.2  &  6.6  &  2 \\
F0  &  16.1  &  48.4  &  117 \\
G0  &  16.9  &  331.8  &  5524 \\
K0  &  17.2  &  667.3  &  22344 \\
M0  &  17.5  &  1279.3  &  82120 \\
\hline
\end{tabular}
\end{center}
\end{table}

\label{lastpage}


\begin{thebibliography}{}

\bibitem[Alcock et al.\ 2000]{alcock00} Alcock C., et al., 2000,
MNRAS, 316, 729 
\bibitem[Allen 1973]{allen73} Allen C.W., 1973, Astrophysical
Quantities. University of London, Athlone Press, 3rd ed. 
\bibitem[Bonnet-Bidaud et al.\ 1989]{bonn89} Bonnet-Bidaud J.M., Motch
C., Beuermann K., Pakull M.W., Parmar A.N., van der Klis M., 1989,
A\&A, 213, 97 
\bibitem[Callanan et al.\ 1990]{call90} Callanan P.J., Charles P.A.,
van Paradijs J., van der Klis M., Pedersen H., Harlaftis E.T., 1990, A\&A, 240, 346
\bibitem[Crampton et al.\ 1990]{cramp90} Crampton D., Cowley A.P.,
Hutchings J.B., Schmidtke P.C., Thompson I.B., 1990, ApJ, 355, 496 
\bibitem[Edelson \& Krolik 1988]{edel88} Edelson R.A., Krolik J.H.,
1988, ApJ, 333, 646 
\bibitem[Gaskell \& Peterson 1987]{gask87} Gaskell C.M., Peterson
B.M., 1987, ApJS, 65, 1 
\bibitem[Gottlieb, Wright \& Liller 1975]{gott75} Gottlieb E.W.,
Wright E.L., Liller W. 1975, ApJ, 195, L33 
\bibitem[Harlaftis, Charles \& Horne 1997]{harl97} Harlaftis E.T.,
Charles P.A., Horne K., 1997, MNRAS, 285, 673 
\bibitem[Hynes et al.\ 1998]{hynes98} Hynes R.I., et al., 1998, MNRAS,
299, L37 
\bibitem[Ilovaisky et al.\ 1980]{ilov80} Ilovaisky S.A., et al., 1980,
MNRAS, 191, 81, (I80)
\bibitem[Johnston, Bradt \& Doxsey 1979]{john79} Johnston M.D., Bradt
H.V., Doxsey R.E., 1979, ApJ, 233, 514 
\bibitem[Koen 1994]{koen94} Koen C., 1994, MNRAS, 268, 690
\bibitem[Koen 2003]{koen03} Koen C., 2003, MNRAS, in press
\bibitem[Kong et al.\ 2000]{kong00} Kong A.K.H., Homer L., Kuulkers
E., Charles P.A., Smale A.P., 2000, MNRAS, 311, 405
\bibitem[LaSala \& Thorstensen 1985]{lasala85} LaSala J., Thorstensen
J.R., 1985, AJ, 90, 2077 
\bibitem[Leong et al.\ 1971]{leong71} Leong C., Kellog E., Gursky H.,
Tananbaum H., Giacconi R., 1971, ApJ, 170, L67
\bibitem[Lomb 1976]{lomb76} Lomb N.R., 1976, Astrophys. Space Sci., 39, 447
\bibitem[Long, Helfand \& Grabelsky 1981]{long81} Long K.S., Helfand
D.J., Grabelsky D.A., 1981, ApJ, 248, 925 
\bibitem[Markert \& Clark 1975]{mark75} Markert T.H., Clark G.W.,
1975, ApJ, 196, L55 
\bibitem[Matsuoka et al.\ 1984]{matsu84} Matsuoka M., et al., 1984,
ApJ, 283, 774 
\bibitem[Middleditch \& Priedhorsky 1985]{middle85} Middleditch J.,
Priedhorsky W., 1985, IAU Circ., 4060 
\bibitem[Motch et al.\ 1985]{motch85} Motch C., Chevalier C.,
Ilovaisky S.A., Pakull M.W., 1985, Space Sci. Rev., 40, 239 
\bibitem[Motch \& Pakull 1989]{motch89} Motch C., Pakull M.W., 1989,
A\&A, 214, L1 
\bibitem[O'Brien et al.\ 2002]{obrien02} O'Brien K., Horne K., Hynes
R.I., Chen W., Haswell C.A., Still M.D., 2002, MNRAS, 334, 426
\bibitem[O'Donoghue 1995]{odon95} O'Donoghue D., 1995, Baltic Astron.,
4, 519 
\bibitem[Pakull 1978]{pak78} Pakull M.W., 1978, IAU Circ 3313
\bibitem[Pakull \& Swings 1979]{pak79} Pakull M.W., Swings J.P. 1979,
IAU Circ 3318 
\bibitem[Pedersen et al.\ 1982]{peder82} Pedersen H., et al., 1982,
ApJ, 263, 325 
\bibitem[Petro et al.\ 1981]{petro81} Petro L.D., Bradt H.V., Kelley
R.L., Horne K., Gomer R., 1981, ApJ, 251, L7, (P81)
\bibitem[Roberts, Lehar \& Dreher 1987]{rob87} Roberts D.H., Lehar J.,
Dreher J.W., 1987, AJ, 93, 968 
\bibitem[Smale \& Kuulkers 2000]{smale00} Smale A., Kuulkers E., 2000,
ApJ, 528, 702 
\bibitem[Stetson 1987]{stet87} Stetson P.B., 1987, PASP, 99, 191
\bibitem[van Paradijs 1983]{vanP83} van Paradijs J., 1983, in
Accretion-Driven Stellar X-ray Sources, ed. Lewin, van den Heuvel,
189, Cambridge University Press, Cambridge 
\bibitem[Webbink, Rappaport \& Savonije 1983]{webb83} Webbink R.F.,
Rappaport S., Savonije G.J., 1983, ApJ, 270, 678
\bibitem[White 1989]{white89} White N.E., 1989, Astron. Ast. Rev., 1, 85

\end{thebibliography}
\end{document}